
\font\sixrm=cmr6           \font\sevenrm=cmr7          
\font\tenrm=cmr10
\font\sixi=cmmi6                     
\font\teni=cmmi10
\font\sixsy=cmsy6                   
\font\tensy=cmsy10
   \font\sevenit=cmti7         
\font\tenit=cmti10
   \font\sevensl=cmsl8 at 7pt  
\font\tensl=cmsl10
          \font\sevenbf=cmbx7         
\font\tenbf=cmbx10
   \font\seventt=cmtt8 at 7pt  
\font\tentt=cmtt10

\font\twelverm=cmr12             \font\ninerm=cmr9
             \font\ninei=cmmi9
    \font\ninesy=cmsy9
            \font\nineit=cmti9
            \font\ninesl=cmsl10 at 9pt
            \font\ninebf=cmbx9
            \font\ninett=cmtt10 at 9pt

\font\fiverm=cmr5               \font\fourrm=cmr5 at 4pt
\font\fivei=cmmi5               \font\fouri=cmmi5 at 4pt
\font\fivesy=cmsy5              \font\foursy=cmsy5 at 4pt


\font\sevenex=cmex10 at 7pt
\def\tenpoint{%
\def\rm{\fam0\tenrm}%
\def\it{\fam\itfam\tenit}%
\def\sl{\fam\slfam\tensl}%
\def\bf{\fam\bffam\tenbf}%
\def\tt{\fam\ttfam\tentt}%
 \textfont0=\tenrm   \scriptfont0=\sixrm \scriptscriptfont0=\fiverm
 \textfont1=\teni    \scriptfont1=\sixi  \scriptscriptfont1=\fivei
 \textfont2=\tensy   \scriptfont2=\sixsy \scriptscriptfont2=\fivesy
 \textfont3=\tenex   \scriptfont3=\tenex \scriptscriptfont3=\tenex
 \textfont\itfam=\nineit
 \textfont\slfam=\ninesl
 \textfont\bffam=\tenbf
 \textfont\ttfam=\ninett
 \baselineskip=12pt
}
\def\ninepoint{%
\def\rm{\fam0\ninerm}%
\def\it{\fam\itfam\nineit}%
\def\sl{\fam\slfam\ninesl}%
\def\bf{\fam\bffam\ninebf}%
\def\tt{\fam\ttfam\ninett}%
 \textfont0=\ninerm   \scriptfont0=\fiverm \scriptscriptfont0=\fourrm
 \textfont1=\ninei    \scriptfont1=\fivei  \scriptscriptfont1=\fouri
 \textfont2=\ninesy   \scriptfont2=\fivesy \scriptscriptfont2=\foursy
 \textfont3=\sevenex  \scriptfont3=\sevenex 
\scriptscriptfont3=\sevenex
 \textfont\itfam=\sevenit
 \textfont\slfam=\sevensl
 \textfont\bffam=\sevenbf
 \textfont\ttfam=\seventt
 \baselineskip=12pt
}
\font\tensmc=cmcsc10

\font\sixteeni=cmmi12 scaled\magstep1


\hsize     = 128mm
\vsize     = 190mm
\topskip   =  12pt
\leftskip  =  0mm
\parskip   =   0pt
\parindent =   4mm
\hoffset = 18mm
\voffset = 26mm

\def\Raggedright{%
 \leftskip=0pt
 \rightskip=0pt plus \hsize
 \spaceskip=.3333em
 \xspaceskip=.5em}

\def\Fullout{
 \leftskip=0pt
 \rightskip=0pt
 \spaceskip=0pt
 \xspaceskip=0pt}

\newcount\notenumber

\def\note{\global\advance\notenumber by 1
  \footnote{${^\the\notenumber}$}}

\newcount\authornumber

\def\authn{\global\advance\authornumber by 1\relax}

\newcount\titlerows

\def\titlen{\global\advance\titlerows by 1\relax}

\newcount\firstpageno
\newcount\lastpageno
\newcount\tabno

\newcount\figno

\def\rheadl{0}
\def\rheadr{0}
\def\studia{\sevenrm\frenchspacing Studia geoph. et geod. 42 (1998)}
\def\geos{\sevenrm\frenchspacing\copyright\ 1998 StudiaGeo s.r.o.,
Prague}
\def\blank{\quad\hfil\quad}
\def\revision{\bigskip
\line{\nineit Manuscript received: 29th January, 1998;\hfil
              Revisions accepted: 15th May 1998}}


\def\ct#1\par{
 \twelverm\baselineskip=14pt
 \centerline{\uppercase{#1}}
 \titlen
 \ifnum\titlerows=1
 {\global\def\rhdr{\nineit #1}}{\global\def\rheadr{\nineit #1}}
 \else{\global\def\rheadr{\rhdr{\nineit\ ...}}}\fi}

\def\ca#1\par{
 \medskip
 \tensmc\baselineskip=12pt
 \centerline{#1}
 \authn
 \ifnum\authornumber=1
 {\global\def\rhdl{\nineit #1}}{\global\def\rheadl{\nineit #1}}
 \else
  \ifnum\authornumber=2{\global\def\rheadl{\rhdl{\nineit\ and\ #1}}}
  \else {\global\def\rheadl{\rhdl {\nineit\ et al.}}}\fi
 \fi}

\def\aa#1#2\par{
 \nineit\baselineskip=12pt
 \centerline{#1\note{\ninerm Address:\ #2}}
 }

\def\abstract#1\par{
\bigskip
 {\ninerm S u m m a r y :\ }
 \ninepoint\it\baselineskip=12pt
 #1\par}

\def\keywords#1\par{
\bigskip
\ninepoint\rm {K e y\quad w o r d s :\ }
 #1\par}

\def\ha#1\par{
 \goodbreak
 \Raggedright
 \tenrm\baselineskip=15pt
 \bigskip
 \centerline{\uppercase{#1}}
 \medskip}

\def\hb#1\par{
 \goodbreak
 \Raggedright
 \tenrm\baselineskip=15pt
 \bigskip
 \centerline{#1}
 \medskip}

\def\tx{
 \Fullout
 \baselineskip12pt
 \tenpoint\rm }

\def\bb#1\par{
  \hangindent=8mm
  \hangafter=1
  \frenchspacing
   \Fullout
 \ninepoint\rm
 \baselineskip=10pt
 \medskip
   \noindent{#1}\par}

\def\ref{\bigskip 
 \centerline{\nineit References}}

\def\tab#1\par{\bigskip 
{\global\advance\tabno by 1 \relax}
\ninepoint\rm
\baselineskip=10pt
\goodbreak
\noindent
{Table\ \number\tabno.\ #1}\smallskip}

\def\fig#1\par{
{\global\advance\figno by 1 \relax}
\ninepoint\rm
\baselineskip=10pt
\smallskip
\noindent
{Fig.\ \number\figno.\  #1}\bigskip}

\def\acknowledge#1\par{
\ninepoint\rm
\baselineskip=10pt
\medskip
{\it Acknowledgements:\ }#1\par}


\def\sethead{
\headline{\ifnum\pageno=\firstpageno\ \hfil\else
\ifodd\pageno\centerline{\rheadr}\else
\ifnum\pageno=\lastpageno\centerline{\rheadl :\ \rheadr}\else
\centerline{\rheadl}\fi
\fi\fi}}

\def\setfoot{
\footline{\ifodd\pageno\studia\ifnum
\firstpageno=\pageno,\ \number\firstpageno-\number\lastpageno\else
\relax\fi\hfil{\twelverm\folio}\else{\twelverm\folio}\hfil\studia\ifnu
\firstpageno=\pageno,\ \number\firstpageno-\number\lastpageno\else
\relax\fi\fi}}

\def\setfootm{
\advance\vsize by -1\baselineskip
\def\makefootline{\lineskip=12pt\baselineskip=10pt
 \vbox{\raggedright\noindent\ifodd\pageno\studia\ifnum
\firstpageno=\pageno,\ \number\firstpageno-\number\lastpageno\else
\relax\fi\hfil{\twelverm\folio}\else{\twelverm\folio}\hfil
\studia\ifnum
\firstpageno=\pageno,\ \number\firstpageno-\number\lastpageno\else
\relax\fi\fi\ifnum\firstpageno=\pageno\break\geos\hfil
\else\break\line{\blank}\fi
}}}

\firstpageno=1
\lastpageno=6
\nopagenumbers
\pageno=\firstpageno

\textfont1=\sixteeni
\ct Mean Field Dynamos\par
\ct With Algebraic and Dynamic $\alpha$--Quenchings\par
\textfont1=\teni
\ca A. Tworkowski\par
\aa{Math. Res. Centre, Queen Mary and Westfield College, London, UK}
   {Mile End Road, London E1 4NS (A.S.Tworkowski@qmw.ac.uk)}\par
\ca E. Covas, R. Tavakol\par
\aa{Astronomy Unit, Queen Mary and Westfield College, London, UK}
{M.E. Rd., London E1 4NS
(E.O.Covas@qmw.ac.uk; reza@maths.qmw.ac.uk)}
\par
\ca and A. Brandenburg\par
\aa{Dept. of Mathem., University of Newcastle upon Tyne, UK}
{NE1 7RU, Newcastle upon Tyne (Axel.Brandenburg@ncl.ac.uk)}\par
\sethead
\setfootm
\abstract  Calculations  for  mean field dynamo models (in both full
spheres and spherical  shells),  with  both  algebraic  and  dynamic
$\alpha$--quenchings,  show  qualitative  as  well  as  quantitative
differences and similarities in the  dynamical  behaviour  of  these
models.    We  summarise  and  enhance  recent  results  with  extra
examples.

Overall, the effect of  using  a  dynamic  $\alpha$  appears  to  be
complicated  and  is  affected  by  the  region  of  parameter space
examined.

\keywords Mean field dynamo, $\alpha$ quenching

\ha 1. Introduction\par

\tx In most studies of axisymmetric mean field  dynamo  models  (see
for  example,  {\it  Tavakol  et  al.,  1995}),  the nonlinearity is
introduced through an algebraic form of  $\alpha$--quenching.   Such
models have produced a number of modes of behaviour in spherical and
spherical shell dynamo models, including periodic, quasiperiodic and
chaotic  solutions.   In  addition  it  has recently been shown that
spherical shell models with this type of quenching  are  capable  of
producing   various  forms  of  intermittent  type  behaviour  ({\it
Tworkowski  et  al.,  1998}),  which  could  be  of   relevance   in
understanding  some  of  the  intermediate  time  scale  variability
observed in the output of the Sun and stars.

Since    such    algebraic    forms   of   $\alpha$--quenching   act
instantaneously, it is possible that this may have a bearing on  the
occurrence of the more complicated modes of behaviour, such as chaos
and  intermittency,  observed  in  these  models.  Here using recent
results as well as new ones, we  make  a  brief  comparison  between
models  with  dynamic  and algebraic $\alpha$--quenching and examine
whether features, such as chaotic and  intermittent-type  behaviour,
survive as $\alpha$--quenching is made dynamic.

\ha 2. The model\par

\tx  The  standard  mean field dynamo equation (cf.  {\it Krause and
R\"adler, 1980}) is
$$
{{\partial {\bf B}}\over{\partial t}}=\nabla \times \left( \bf{u}
\times{\bf B}+\alpha{\bf B} - \eta_t \nabla \times{\bf B} \right),
\eqno{(1)}
$$
where as usual, the magnitudes of the $\alpha$ and $\omega$  effects
are given by the dynamo parameters $C_{\alpha}$ and $C_{\omega}$.

In  the  algebraic case we use a purely hydrodynamical $\alpha$ with
$\alpha=\alpha_h$, for which we take the usual  functional  form  of
$\alpha$--quenching
$$
\alpha_h={{\alpha_0 \cos \theta}\over{1+ \bf{B}^2}}.\eqno{(2)}
$$
For  the dynamical case, following to {\it Zeldovich et al.  (1983)}
and {\it Kleeorin and Ruzmaikin (1982)} (see also {\it  Kleeorin  et
al.,  1995}),  $\alpha$  can  be  divided  into a hydrodynamic and a
magnetic part,
$$
\alpha=\alpha_h+\alpha_m,\eqno{(3)}
$$
where $\alpha_h$ is given by Eq.~(2) and the magnetic part satisfies
an  explicitly  time  dependent  diffusion  type  equation  with   a
nonlinear forcing in the form
$$
{{\partial\alpha_m}\over{\partial t}}={1\over{\mu_0\rho}}
\left({\bf J}\cdot{\bf B}-{{\alpha{\bf B}^2}\over{\eta_t}}\right)
+\nu_\alpha\nabla^2\alpha_m,\eqno{(4)}
$$
(see {\it Covas et al., 1997a,b; 1998a}).  This equation differs from
that  used  by  {\it  Kleeorin  et  al.  (1995)} in that we take the
damping term to be that of the form $\nu_\alpha  \nabla^2  \alpha_m$
instead  of  $-\alpha_m/T$,  where  $T$  is  some damping time.  Our
approach is motivated by stability considerations.

We solved the above equations using spherical polar coordinates.  We
shall consider  both  axisymmetric  spherical  and  spherical  shell
models, where the outer boundary in both cases is denoted by $R$ and
in  the  spherical  shell models, the fractional radius of the inner
boundary of the shell is denoted by $r_0$.  We discuss the behaviour
of the dynamos by calculating the total magnetic energy, $E$,  which
is  split  into  two parts, $E=E^{(A)}+E^{(S)}$, where $E^{(A)}$ and
$E^{(S)}$ are respectively the energies of the field whose  toroidal
component  is  antisymmetric  and  symmetric about the equator.  The
overall parity $P$ given by $P=(E^{(S)}-E^{(A)})/E$, with $P=-1, +1$
denoting   the    antisymmetric    (dipole-like)    and    symmetric
(quadrupole-like) pure parity solutions respectively.

For $\bf{B}$ we assume vacuum boundary conditions and for $\alpha_m$
we use $\alpha_m=0$ on the inner and outer boundary.

\ha 3. Results\par

\tx For the  the  algebraic  $\alpha$--quenching  model,  we  solved
equations  (1)  and  (2),  whilst  for  the dynamical case we solved
equations (1) and (4).  Clearly our conclusions are subject  to  the
finiteness  of  the resolution of parameter space that we chose.  We
also point out that the parameters $C_{\alpha}$ and $C_{\omega}$  do
not   play  identical  roles  in  these  models  and  therefore  the
comparison  of  the   behaviours   in   the   two   cases   is   not
straightforward.

\hb 3.1~~S p h e r i c a l~~~d y n a m o~~~m o d e l s\par

\tx  Taking  $C_{\omega}=-10^4$,  which  is a typical value that has
been used in other dynamo studies, for example {\it Tavakol  et  al.
(1995)} (and references therein), we found that as $C_\alpha$ became
large,  the  algebraic  model  tended  to an antisymmetric (dipolar)
parity  whilst  the  dynamic  case  tended   towards   a   symmetric
(quadrupolar)  one.   Overall,  no chaos was observed in either case
although the dynamic case did  show  evidence  of  more  complicated
behaviour  in  its  transitions from one parity form to another.  We
also  wished  to  study  the  allowed  forms  of  behaviour  in  the
supercritical  regimes,  so  we chose $C_{\omega}$ values which were
effectively the highest values numerically allowed by our code,  and
these  turned  out  to  be  $-10^4$  and  $-10^5$ in the dynamic and
algebraic cases respectively.  For the  algebraic  model,  we  found
chaotic  behaviour at large $C_\alpha$ value when $C_{\omega}=-10^5$
was used.

\hb 3.2~~S p h e r i c a l~~~s h e l l~~~d y n a m o~~~m o d e l s

\tx Using  $C_{\omega}=-10^4$,  we  found  that  for  the  dynamical
$\alpha$--quenching   case,   the   behaviour   observed  for  large
$C_{\alpha}$ values depended on  the  shell  thickness.   For  thick
shells ($r_0=0.2$) we observed symmetric parity solutions whilst for
medium   ($r_0=0.5$)   and   thin   ($r_0=0.7$)   shells   we  found
antisymmetric asymptotic parity solutions.  In particular,  for  the
medium  shell  we  observed multiple attractors at $C_\alpha \approx
15$, that is, we observed  the  occurrence  of  different  dynamical
solutions  possessing  different  mean energies at the same value of
$C_\alpha$ but having different  initial  parities.   Also,  chaotic
behaviour  was  observed  in  the  thin  shell  at $C_\alpha$ values
around $35$.  However, the algebraic $\alpha$--quenching  model  did
not  exhibit  any  chaotic  behaviour  nor  the presence of multiple
attractors at this value of $C_\omega$.   The  asymptotic  behaviour
obtained  for the algebraic case was a symmetric parity solution, an
oscillatory mixed parity solution which varied between symmetric and
antisymmetric values but was periodic and  a  non-oscillatory  mixed
parity  solution for the thick, medium and thin shells respectively.
However,  when  $C_\omega$  was  set  to  $-10^5$,   thus   in   the
supercritical  regime  for  this  case, the asymptotic behaviour was
chaotic for all shell thicknesses.

Our results appear  to  indicate  that  dynamic  $\alpha$--quenching
smoothes out chaotic behaviour possibly because of the back reaction
of $\alpha_m$ via diffusion.

\topinsert
\vbox{\includegraphics{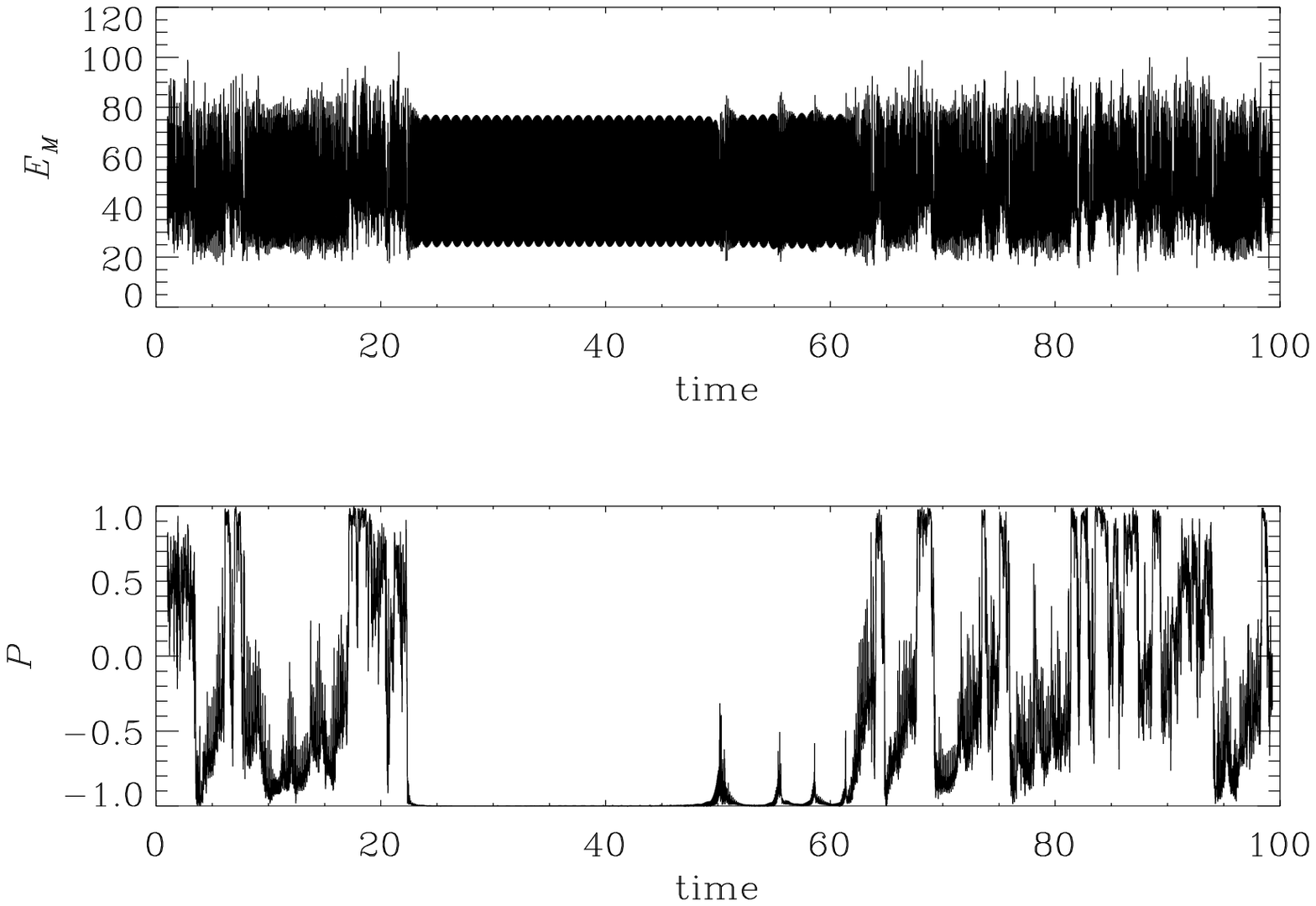}}
\vskip8.7cm
\fig  Total  energy  and  parity  for  the $\alpha$-profile given by
(2).  $C_\alpha=1.94$, $r_0=0.4$, $F=0$\par
\endinsert
\topinsert
\vbox{\includegraphics{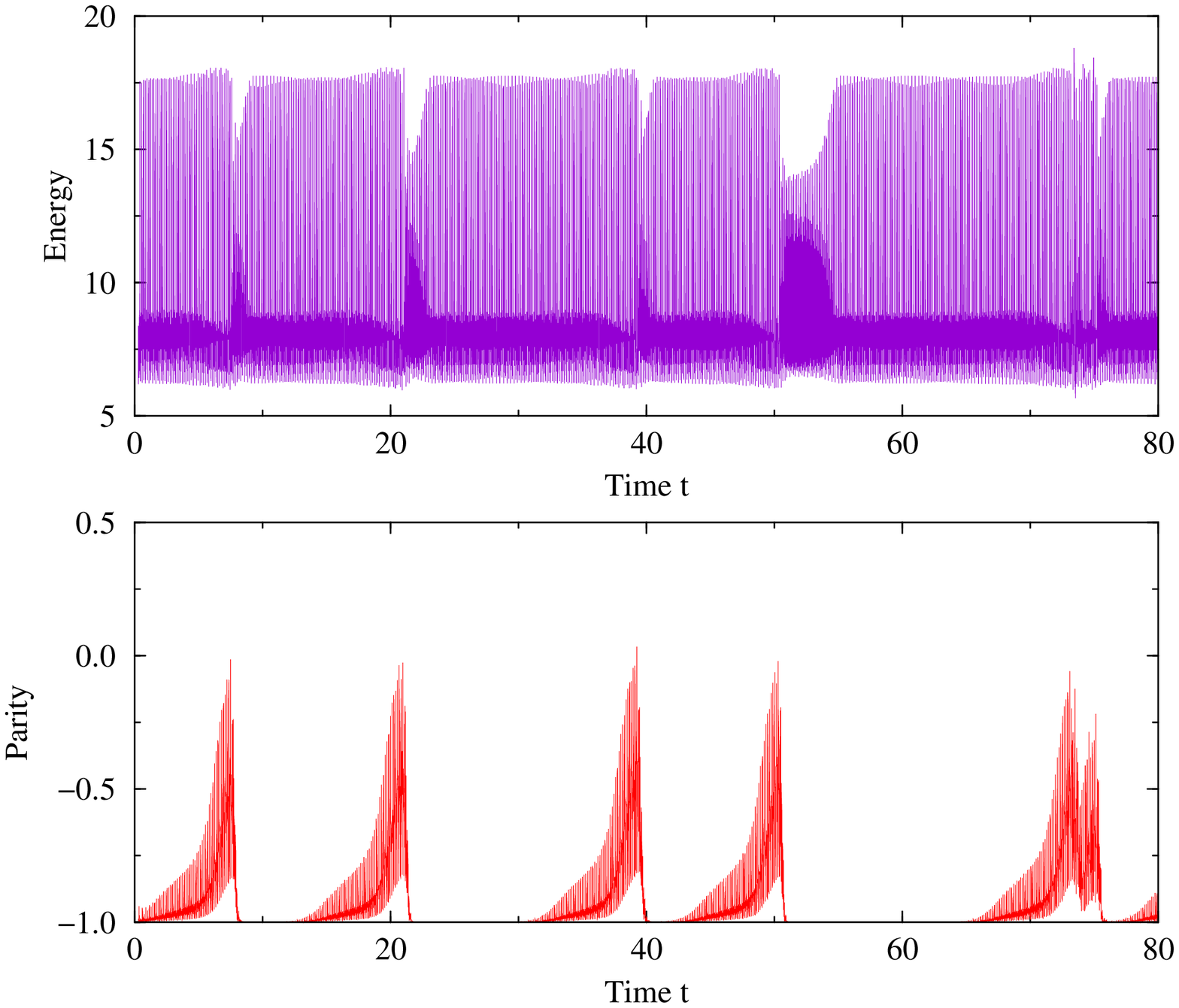}}
\vskip11.3cm
\fig Result for $r_0=0.5$ with the  dynamic  $\alpha$  model  and  a
particular  form  of  $\alpha_h$  (see  text).   The  parameters are
$C_{\alpha}=9.34$, $C_{\omega}=-10^4$ and $F=0$.\par
\endinsert

\ha 4. Intermittency

\tx Spherical shell dynamo models with algebraic $\alpha$--quenching
have  been  shown  to  be  capable  of  producing  various  forms of
intermittent-type of behaviour ({\it Tworkowski et al.,  1998}),  in
which  the  dynamical  modes  of  behaviour for which the statistics
taken over different time intervals are different.  One  example  of
such  behaviour  is  given  in Figure 1 (see {\it Tworkowski et al.,
1998} for other examples).  As can be seen,  for  times
around   $20$  to  $60$ units,  the  parity is  almost  antisymmetric
but that there are also time intervals which are  interrupted  by
excursions  of  the
parity  to  values  well  away from $-1$.  We have also searched for
occurrences of  intermittency  in  the  dynamic  $\alpha$--quenching
model  using the algebraic $\alpha$--quenching given by equation (2)
with no success,  at  least  to  the  resolution  of  our  parameter
regime.   However,  we  did  find  examples  of  intermittency using
another form of  algebraic  quenching,  namely,  that  due  to  {\it
Kitchatinov  (1987)}.   This  form  was  derived  in  the context of
$\Lambda$--quenching  by  rapid  rotation  and  is  essentially   an
interpolation  formula  having the correct asymptotic behaviour.  An
example of the intermittent behaviour found using this form is given
in Figure 2.  Although these two examples  appear  similar  in  that
they  both  have intervals where the parity is nearly antisymmetric,
this does not mean that these two are examples of the same  type  of
intermittency.   To  resolve  this  issue,  one needs to examine the
appropriate theoretical framework for the detailed properties of the
corresponding model such as scaling laws  (see,  for  example,  {\it
Ashwin et al., 1998; Covas et al., 1998b}).

Overall, then, it appears that dynamic  $\alpha$--quenching  appears
to  suppress  the  existence  of  intermittency,  at  least  to  the
resolution of our parameter search.

\ha 5. Conclusion\par

\tx Our studies indicate that in the full sphere  models,  the  main
similarities  are  the presence of similar modes of parity behaviour
and the absence of intermittency, whilst the differences are in  the
details  of the transitions between the different modes of behaviour
and  the  presence   of   chaotic   behaviour   in   the   algebraic
$\alpha$--quenching case.

For the spherical shell models, we observe differences which  depend
on  the  region  of  the  parameter  space  considered.  The dynamic
$\alpha$ models appear to produce more  varied  modes  of  behaviour
for  $C_{\omega}=-10^4$.   However, taking the numerical upper bound
of  $C_{\omega}$  in  each  case  appears  to  indicate   that   the
introduction  of dynamic $\alpha$--quenching drastically reduces the
likelihood of the occurrence of chaotic behaviour and intermittency,
which was observed in  models  with  algebraic  $\alpha$--quenching.
Our  dynamic  $\alpha$--quenching  models  also show multi-attractor
regimes  with  the  possibility  of  the  final  state   sensitivity
(fragility) with respect to small changes in the initial parity.

With   regards   to  the  extra  complexity  introduced  by  a  time
dependent $\alpha$--quen\-ching, our present results show  that  the
behaviour  seen in the solutions is complicated and with the outcome
depending on the region of parameter space considered, rather than a
simplistic decrease or increase in complexity.

\acknowledge EC is supported by grant BD /  5708  /  95  --  Program
PRAXIS XXI, from JNICT -- Portugal.  RT benefited from SERC UK Grant
No.  L39094.  This research also benefited from the EC Human Capital
and   Mobility   (Networks)  grant  ``Late  type  stars:   activity,
magnetism, turbulence'' No.  ERBCHRXCT940483.

\revision
\ref

\bb   Ashwin,   P.,   Covas,   E.    and  Tavakol,  R.,  1998:  
Transverse instability for non-normal parameters. {\it
Nonlinearity}, in press

\bb Covas, E., Tworkowski, A., Brandenburg,  A.   and  Tavakol,  R.,
1997a: Dynamos with different formulations of a dynamic $\alpha$--effect.
{\it Astron.  and Astrophys.} {\bf 317}, 610

\bb Covas, E., Tworkowski, A., Tavakol,  R.   and  Brandenburg,  A.,
1997b: Robustness of truncated $\alpha \omega$ dynamos with a
dynamic $\alpha$. {\it Solar Physics} {\bf 172}, 3

\bb Covas.  E., Tavakol, R., Tworkowski, A.   and  Brandenburg,  A.,
1998a:  Axisymmetric mean field dynamos with dynamic and algebraic 
$\alpha$--quenchings. {\it Astron.  and Astrophys.} {\bf 329}, 350

\bb Covas.  E., Tavakol, R., Ashwin, P., Tworkowski, A.   and  Brooke,
J. M.,
1998b: In--out intermittency in PDE and ODE models of axisymmetric mean-field 
dynamos. Submitted to {\it Phys. Rev. Lett.}

\bb  Kitchatinov,  L.L.,  1987:  A mechanism for differential
rotation based on angular momentum transport by compressible
convection.  {\it  Geophys.   Astrophys.  Fluid
Dyn.} {\bf 38}, 273

\bb  Kleeorin,  N.    I.    and   Ruzmaikin,   A.A.,   1982: Dynamics of
the average turbulent helicity in a magnetic field.   {\it
Magnetohydrodynamica} {\bf N2}, 17

\bb  Kleeorin,  N.   I.,  Rogachevskii, I.  and Ruzmaikin, A., 1995:
Magnitude of the dynamo--generated magnetic field in solar--type
convective zones. {\it Astron.  and Astrophys.} {\bf 297}, 159

\bb  Krause,  F.   and  R\"adler,  K.-H.,  1980:   {\it   Mean-Field
Magnetohydrodynamics and Dynamo Theory}, Pergamon Press, Oxford

\bb  Tavakol,  R.K.,  Tworkowski,  A.  S., Brandenburg, A., Moss, D.
and Tuominen, I., 1995:  Structural stability of axisymmetric
dynamo models. {\it Astron.  and Astrophys.} {\bf 296}, 269

\bb Tworkowski, A., Tavakol, R., Brandenburg, A., Brooke, J. M.,  Moss,
D.  and Tuominen, I., 1998:  Intermittent behaviour in 
axisymmetric mean-field dynamo models in spherical shells.
{\it Mon.  Not.  Roy.  Astr.  Soc.}, in
press

\bb  Zeldovich,  Ya.B.,  Ruzmaikin,  A.A.  and Sokoloff, D.D., 1983:
{\it Magnetic Fields in Astrophysics}, Gordon and Breach, New York

\bye